\documentclass[aps,twocolumn]{revtex4-1} 
\usepackage{amsmath}
\usepackage{amssymb}
\usepackage{bm}
\usepackage{color}
\usepackage[dvipdfmx]{graphicx}
\newcommand{\tfix}[1]{{\textcolor{black}{#1}}}

\def\ket#1{\mathinner{|{#1}\rangle}}
\def\braket#1{\mathinner{\langle{#1}\rangle}}

\begin{document}
  \title{Correlation effects on the magnetization process of the Kitaev model}
  \author{Kota Ido and Takahiro Misawa}
  \affiliation{Institute for Solid State Physics, University of Tokyo, 5-1-5 Kashiwanoha, Kashiwa, Chiba 277-8581, Japan}

\begin{abstract}
  By using the variational Monte Carlo method, we study the magnetization process of 
  the Kitaev honeycomb model in a magnetic field.
  Our trial wavefunction is a generalized Bardeen-Cooper-Schrieffer wave function 
  with the Jastrow correlation factor, which exactly describes 
  the ground state of the Kitaev model at zero magnetic field using 
  the Jordan-Wigner (JW) transformation.
  We find that two phase transitions occur for the antiferromagnetic Kitaev coupling, 
  while only one phase transition occurs for the ferromagnetic Kitaev coupling.
For the antiferromagnetic Kitaev coupling,
  we also find that the topology of the momentum distribution of the 
JW fermions 
  changes at the transition point from the Kitaev spin 
  liquid to an intermediate state.
  Our numerical results indicate that the intermediate state between the 
Kitaev spin liquid and the fully polarized phases stably 
exists in the bulk system on two dimensions 
for the antiferromagnetic Kitaev 
  coupling against many-body correlations.
\end{abstract}

\maketitle
%
\section{Introduction}
%
Quantum many-body interactions often
induce  
fractionalizations of the internal
degrees of freedom of electrons, resulting in exotic elementary excitations.
For example, in the one-dimensional Hubbard model, 
it is known that the spin degrees of freedom separate
from the charge degrees of freedom (spin-charge separation) \cite{Essler2005}.
Another interesting possibility is the emergence of 
Majorana fermions, which can be regarded as
a fractionalization of the spin degrees of freedom.
Although 
the Majorana fermions appear at the edges/surfaces of the
topological superconductors\cite{Read2000,Fu2008}, 
this has not yet been fully settled experimentally\cite{Mourik2012,Nadj-Perge2014,Albrecht2016,He2017,Zhang2018,Zhang2019,Machida2019,Wang2018,Chen2018}.

Kitaev has proposed another way to realize Majorana fermions: 
they appear
as elementary excitations 
in a quantum spin model on a honeycomb lattice with bond-dependent Ising interactions\cite{Kitaev2006}. 
This is often called the Kitaev model.
It has been shown that its ground state is a quantum spin liquid that 
can be represented by non-interacting Majorana fermions\cite{Kitaev2006,Feng2007,Chen2008}.
Thus, the Kitaev model offers an ideal platform for investigating Majorana fermions.
This model may be realizable by utilizing the interaction between electrons and spin-orbital couplings 
\cite{Jackeli2009,Chaloupka2010}. 
In fact, the signature of the Kitaev spin liquid has actually been observed
in iridium\cite{Gretarsson2013,Modic2014,HwanChun2015,Kitagawa2018,Revelli2019} and in ruthenium oxides\cite{Sandilands2015,Nasu2016,Banerjee2016,Zheng2017,Banerjee2017,Kasahara2018}.
Recently, the half-quantized thermal Hall conductivity,
which is a direct evidence for Majorana fermions\cite{Nomura2012,Sumiyoshi2013},
has been reported in $\alpha$-RuCl$_{3}$ in a tilted magnetic field\cite{Kasahara2018}.

A magnetic field is known to induce interactions between Majorana fermions in the Kitaev model\cite{Kitaev2006}.
Currently, one of the hottest issues in Kitaev materials is
how the magnetic fields change the nature of non-interacting Majorana fermions.
The Kitaev model in a magnetic field has therefore been studied intensively to clarify the nature of interacting Majorana fermions\cite{Jiang2011,Janssen2016,Yadav2016,Liu2018,Joshi2018,Gohlke2018a,McClarty2018,Zhu2018,Fu2018,Gohlke2018,Nasu2018,Liang2018,Hickey2019}.
Recently, an intermediate gapless state 
has been reported between the Kitaev spin liquid and the polarized phase 
for the antiferromagnetic Kitaev coupling 
in an applied magnetic field\cite{Zhu2018,Fu2018,Gohlke2018,Nasu2018,Liang2018}.
The appearance of the intermediate phase was reported for a [111] magnetic field using the exact diagonalization (ED) method and the density-matrix renormalization group (DMRG) method\cite{Zhu2018,Fu2018,Gohlke2018,Hickey2019}.
A similar intermediate phase was found for the [001] case, mainly using the mean-field (MF) approximations\cite{Nasu2018,Liang2018}.
However, applications of ED and DMRG are limited to 
small or quasi-one-dimensional systems, and
the range of applicability of the MF approximation is not clear.
Accurate analyses beyond the MF approximation for larger
system sizes in two dimensions are thus necessary 
to investigate the emergence of the intermediate 
phase in a magnetic field.

In this paper, we use the variational Monte Carlo (VMC) method \cite{Gros1989,becca2017quantum} to investigate whether the intermediate state exists as 
the ground state of the Kitaev model in two dimensions in a [001] magnetic field. 
The VMC method enables us to treat 
strongly correlated electron systems with high accuracy\cite{LeBlanc2015,Tahara2008a,Tocchio2008,Misawa2014a,Zhao2017,becca2017quantum}.
We used a generalized Bardeen-Cooper-Schrieffer (BCS) trial wavefunction with Jastrow correlation factor.
Our benchmark comparisons with the ED method for a small system show that this trial wavefunction reproduces the magnetization process well.
In applications to large systems with antiferromagnetic Kitaev coupling, 
we find a double-peaked structure in the magnetic susceptibility,
which signals the existence of the intermediate phase.
\tfix{In the intermediate, we find that the local $Z_{2}$ gauge field takes non-integer values, 
which is consistent with the spin liquid phase with fluctuating flux for [111] case \cite{Hickey2019}. }
We also find that the momentum distribution of the complex fermions changes 
topologically at the first transition point, suggesting that 
this is a continuous topological phase transition resembling the Lifshitz transition. 
These properties are not found in the ferromagnetic Kiteav model.
From these results, we conclude that for the antiferromagnetic Kitaev coupling, the intermediate state 
is stable against many-body correlations beyond the MF approximation.

%
\section{Model and Method}
%
The Kitaev honeycomb model in a [001] magnetic field is defined by 
\begin{eqnarray}
  \mathcal{H} =&& \sum_{\gamma=x,y,z} \sum_{\braket{I,J} \in \gamma-{\rm bond}} K_\gamma S^\gamma_I S^\gamma_J - h\sum_{I} S^z_I. \label{Kitaev}
\end{eqnarray}
Here, $I=(i,\alpha)$ is a site index, defined as 
unit-cell index $i$ with intra-cell degrees of freedom $\alpha=A, B$. 
We consider only the isotropic coupling $K=K_x=K_y=K_z$ with $|K| = 1$.
We solve the fermionized Kitaev model  
by using the Jordan-Wigner (JW) transformation, 
$S^+_ I = \prod_{J < I}\left( -2S^z_J \right) c^\dagger_I$ and $S^z_I = (n_I-\frac{1}{2})$, 
where $c^\dagger_{I}$ ($c_{I}$) 
is the creation (annihilation) operator for a JW electron on the site $I$, and $n_{I} = c^\dagger_{I}c_{I}$.
We assume that $z$-bond is parallel to $y$-axis.
\tfix{To directly compare the ED results for the conventional spin-1/2 basis, we employ the open boundary conditions in benchmarks, 
in which the boundary terms exactly vanish\cite{Chen2008,Feng2007}. 
For larger system sizes, because the effects of the boundary terms become negligibly small, 
we take periodic boundary conditions, which is employed in the previous study\cite{Nasu2018}.}

To simulate magnetization process of the Kitaev spin liquid, 
we used the VMC method which provides us to 
obtain ground states in quantum many-body systems\cite{becca2017quantum}.
As a trial wavefunction for the VMC method, 
we adopted the generalized BCS 
wave function $\ket{\phi}$ with the many-body correlation factor $\mathcal{P}$: 
$\ket{\psi_{\rm JW}} = \mathcal{P} \ket{\phi}$.
Here $\mathcal{P}=\exp \left( \sum_{I,J}v_{IJ} n_{I}n_{J}\right)$ and
 $\ket{\phi} = \exp \left( \sum_{I,J}^{N_s} f_{IJ} c_{I}^\dagger c_{J}^\dagger \right) \ket{0}$, 
where $N_s=2\times L\times L$ is the system size. 
In this study, we treated $v_{IJ}$ and $f_{IJ}$ as variational parameters, and we optimized them simultaneously 
by using the stochastic reconfiguration method~\cite{Sorella2001}.
We performed VMC simulations in the grand canonical ensemble, because 
the transformed Hamiltonian includes BCS terms 
that do not conserve the total number of JW fermions\cite{Feng2007,Chen2008}.
By using the wavefunction $\ket{\psi_{\rm JW}}$, we can exactly represent the ground state of 
the Kitaev model at zero magnetic field\cite{Feng2007,Chen2008}. We can also 
examine the ground state in a magnetic field 
beyond the 
MF approximation, 
thanks to the many-body correlation factor. 
In the actual calculations, we 
ignore the boundary terms involved with the string operator when we consider periodic systems, 
because we expect these effects to vanish in the thermodynamic limit.
We performed the calculations mainly for $K>0$, because 
an antiferromagnetic $K$ induces the 
intermediate state 
reported in the previous studies\cite{Zhu2018,Fu2018,Gohlke2018,Nasu2018,Liang2018}.
Although it has been pointed out that presently known Kitaev-like materials have 
the ferromagnetic Kitaev coupling\cite{Yamaji2014,Kim2015,Yadav2016,Katukuri2016,Winter2016,Jang2019}, 
recent first-principles studies have proposed possible Kitaev-like materials with dominant antiferromagnetic $K$ values, 
such as $f$-electron based magnets\cite{Jang2019} and polar spin-orbit Mott insulators\cite{Sugita2019}.

\begin{figure}[tbp]
  \begin{center}
   \includegraphics[width=80mm]{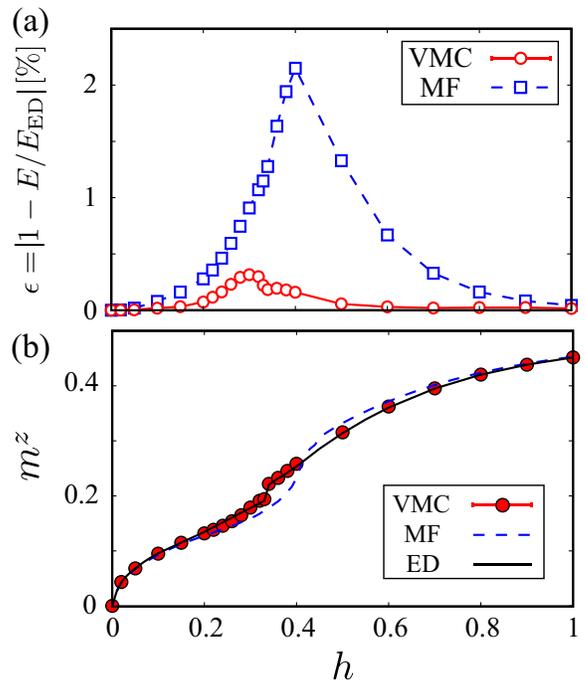}
  \end{center}
  \caption{(Color online) Benchmarks for magnetization process of an antiferromagnetic Kitaev model with $N_s=2\times4\times 3$.
  $h$ is the strength of the magnetic field. 
  The relative error $\epsilon = |1-E/E_{\rm ED}|$ and the magnetic moment $m^z$ are shown in panels (a) and (b), respectively.
  Here $E_{\rm ED}$ is the exact energy of the ground state obtained by ED.
  In panel (a) the red solid line with circles and the blue dashed line with squares represent the results obtained by VMC and MF, respectively. 
  The black line in panel (b) represents the ED result. The types of simulations are identified in the inset in panel (b).
  The statistical errors arising from the Monte Carlo sampling are smaller than the symbol sizes.
  }
  \label{benchmark}
\end{figure} 

%
\section{Results}
%
\subsection{Benchmarks}
First, to check the accuracy of our trial wavefunction, 
we performed benchmark calculations in a $4\times3$ unit cell system with 
the open-open boundary conditions.
Figures \ref{benchmark}(a) and (b) show that the magnetic-field dependence 
of the relative error in the ground-state energy $E=\braket{\mathcal{H}}$ 
and the magnetic moment $m^z = 1/N_s \sum_i \braket{S^z_i}$, respectively.
The MF result obtained by using only $\ket{\phi}$ has good accuracy 
around $h=0$ and $1$, because it becomes the exact ground state at $h=0$
and as $h\rightarrow \infty$.
The accuracy of the MF wavefunction decreases in the intermediate region, 
with the maximum error reaching about 2\% at $h\sim0.4$. 
Because of this poor accuracy in the energy,
the magnetization deviates significantly
from the ED result in the intermediate region.
Compared with the MF result, 
the VMC method by using $\ket{\psi_{\rm JW}}$ improves the accuracy 
for all magnetic fields,
with a maximum error $\sim$ 0.3\%\tfix{, which is much smaller than that in the previous VMC studies by using the wavefunction with different basis\cite{Kurita2015,Jiang2019}}.
We also confirm 
that
our trial wavefunction $\ket{\psi_{\rm JW}}$ reproduces the exact magnetization process well.

\begin{figure}[tbp]
  \begin{center}
   \includegraphics[width=80mm]{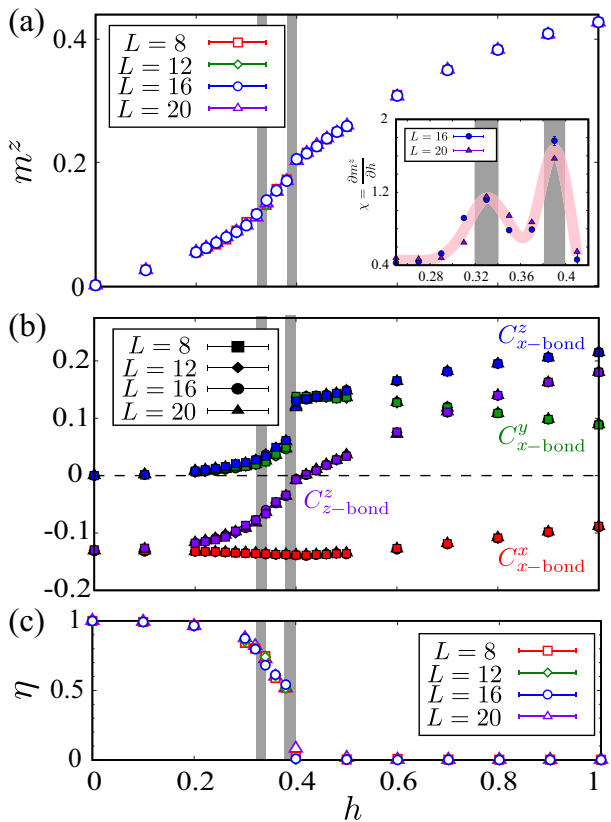}
  \end{center}
  \caption{(Color online) Magnetization process of the antiferromagnetic Kitaev model for $N_s=2\times L \times L$. 
  The system sizes are indicated in the legend.
  The gray bars show the transition points predicted from the peak structures in the susceptibility data.
  (a) The $h$-dependence of the local magnetic moment $m^z$.
  The susceptibility $\chi$ is plotted in the inset, where the 
  wide pink curve is just a guide for the eye.
  (b) The $h$-dependence of the spin-$\lambda$ correlations between the nearest-neighbor sites on $\gamma$-bond $C_{\gamma-{\rm bond}}^{\lambda}$.
  (c) The $h$-dependence of the absolute value of the local ${\rm Z}_2$ gauge field defined on $z$-bond\cite{z2_note}.
  }
  \label{mag}
\end{figure} 

\subsection{Magnetization process}
Next, we analyze the field-dependence of the magnetization and 
magnetic correlations 
in the Kitaev model for larger system sizes, using antiperiodic-periodic boundary conditions.
To reduce the numerical costs, we imposed a $1 \times 1$ sublattice structure on the trial wavefunction.
\tfix{Our simulated systems satisfy the closed shell condition in the noninteracting limit for stable optimization.}
We note that long-period ordered states---such as incommensurate magnetic orders that might become ground states---
cannot be represented by our wavefunction.
To investigate possible phase transitions in a magnetic field, 
we evaluated the magnetic moment $m^z$ and 
the correlation functions between the nearest-neighbor spins of the $\lambda$ component 
on $\gamma$-bond $C^\lambda_{\gamma\rm-bond} = 1/ N_\gamma \sum_{\braket{I,J} \in \gamma{\rm -bond}}\braket{S^\lambda_I S^\lambda_J}$, 
where $N_\gamma$ means the number of the $\gamma$ bonds in the system.

Figure \ref{mag} (a) shows the $h$-dependence of $m^z$. 
The magnetization curve exhibits the signatures of two phase transitions:
A discontinuous phase transition occurs at $h_{c2}\sim0.39$, as evidenced by the jump in magnetization.
Below $h_{c2}$, the slope of the magnetization changes
around $h_{c1}\sim0.33$, which signals a possible continuous phase transition. 
These critical fields are weaker than those of the previous MF results\cite{Nasu2018,Liang2018}.
To see the change in the slope of the magnetization more directly, 
we calculated the field dependence of the magnetic 
susceptibility $\chi=dm^z/dh$, which is plotted in the inset in Fig. \ref{mag} (a).
Around $h\sim0.33$,
there is a broad peak-structure in the magnetic susceptibility, which is the signature of a continuous phase transition.
These results support the existence of the intermediate state pointed by the previous MF studies\cite{Nasu2018,Liang2018}. 
Note that such a double-peaked structure is not found in small systems ($L\leq8$), \tfix{which is consistent with the ED result for 24 sites\cite{Nasu2018}.}
\tfix{This means that} the large-scale calculations are essential for obtaining the phase diagram accurately in a magnetic field.

\begin{figure*}[t]
  \begin{center}
   \includegraphics[width=160mm]{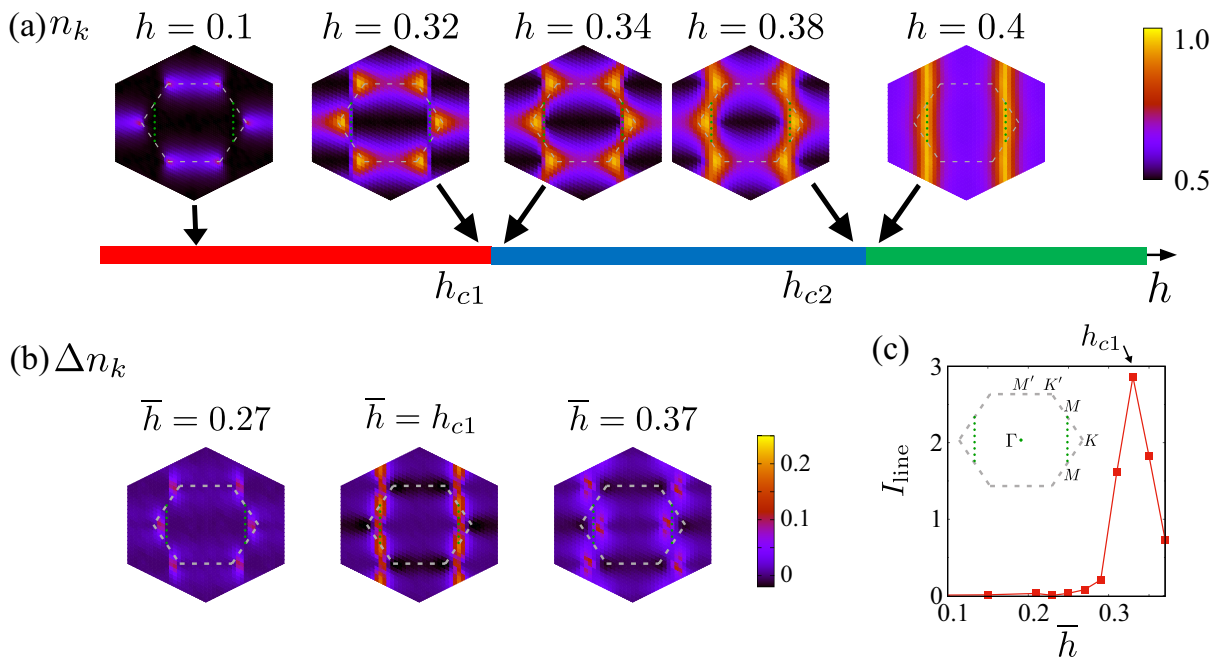}
  \end{center}
  \caption{(Color online) (a) Momentum distribution $n_{\bm{k}}$ in the antiferromagnetic Kitaev model for $L=20$.
  The gray dashed and green dotted lines are the boundary of the first BZ and the M-M line, respectively.
  The color scale represents the value of $n_{\bm{k}}$. 
  Special points in the BZ are denoted in the inset of the panel (c).
  (b) Difference in the momentum distributions $\Delta n_{\bm{k}}(\overline{h})=n_{\bm{k}}(h=\overline{h}+0.1)-n_{\bm{k}}(h=\overline{h}-0.1)$ for $\overline{h}=0.27, h_{c1}$, and $0.37$.
  The color scale represents the value of $\Delta n_{\bm{k}}$. 
  (c) Integrated value of $\Delta n_{\bm{k}}$ around the M-M line $I_{\rm line}(\overline{h}) =\sum_{\bm{k} \in \rm M-M\ line} \Delta n_{\bm{k}}(\overline{h})$.
  }
  \label{nk}
\end{figure*}

Figure \ref{mag} (b) shows the $h$-dependence of $C^\lambda_{\gamma\rm-bond}$.
At $h=0$, $C^\lambda_{\gamma -{\rm bond}}$ has the same 
negative value for $\lambda = \gamma$ and $C^\lambda_{\gamma -{\rm bond}}=0$ for $\lambda \neq \gamma$.
With increasing $h$, the $z$-component of the spin correlation $C^z_{\gamma -{\rm bond}}$ is enhanced.
Although the VMC results are almost the same as the previous MF results\cite{Nasu2018}, 
there is a difference in the strength of the magnetic field at which the sign change of $C^z_{z{\rm-bond}}$ occurs.
A previous study using ED and MF pointed out that the sign change of the effective interaction 
on the $z$ bond happens near $h_{c1}$, and thus it would be related 
to the topological phase transition induced by $h$\cite{Nasu2018}.
However, our results show that this occurs above $h_{c2}$, which implies that 
this sign change in the spin correlations 
bears no relation to the 
possible continuous phase transition.

\tfix{To understand the nature of the intermediate phase, we calculate the expectation value of the local ${\rm Z}_2$ gauge field $\eta$ defined on the $z$-bond 
in the fermionized Kitaev model by the JW transformation\cite{Feng2007,Chen2008}, 
which is shown in Fig. \ref{mag}(c). Here $\eta=\frac{1}{N_z}\sum_{\braket{I,J}\in z{\rm -bond}} \left| \braket{c^\dagger_{I}c_{J} + c^\dagger_{I}c^\dagger_{J}} + {\rm c.c.} \right|$. 
This value shoud be 1 for $h=0$. We found that the intermediate phase has intermediate value between 0 and 1. 
This indicates that the ${\rm Z}_2$ value is fluctuating in the intermediate state for [001], 
which has been consistent with the flux fluctuation in the intermediate phase for [111] pointed out in Ref. \cite{Hickey2019}. 
This result implies that the intermediate phase in [001]
direction has the same physical property with
the intermediate phase in [111] direction.}

%
\subsection{Momentum distribution}
%
To obtain further evidence of the phase transition around $h_{c1}$,
we measured the momentum distributions of 
the JW fermions[Fig. \ref{nk} (a)]. 
The momentum distribution $n_{\bm{k} \alpha}$ is 
defined as $n_{\bm{k} \alpha} = \braket{c^\dagger_{\bm{k}\alpha} c_{\bm{k}\alpha}}$.
Since $n_{\bm{k}A} = n_{\bm{k}B}$, we only show the results for $n_{\bm{k}A} = n_{\bm{k}}$.
For $h < h_{c1}$, there is an occupied state (island) with two strong peaks in the first Brilluin zone (BZ). 
The peaks would correspond to the positions of the Dirac cones in the Majorana spectrum, as we discuss below.
Around $h_{c1}$,
we find that the topology of the momentum distribution changes: 
the islands of JW fermions connect to each other around the M point. 
This topological shape is maintained for $h_{c1} < h < h_{c2}$.
The topology of $n_{\bm{k}}$ becomes more one-dimensional like shape with increasing $h$.
Above $h_{c2}$, the JW fermions are distributed over the entire BZ.
We also find that the dependence of $n_{\bm{k}}$ on the $k_y$ direction 
is strongly suppressed compared with the results below $h_{c2}$.
This means that the JW fermions freeze on the $z$-bond, which is consistent with the ground 
state in this region being fully polarized phase along the $z$ direction. 

To clarify what happens at $h_{c1}$ quantitatively, 
in Figs. 3 (b) and (c), respectively, 
we also show the difference between 
the momentum distribution $\Delta n_{\bm{k}}(\overline{h})=n_{\bm{k}}(h=\overline{h}+0.1)-n_{\bm{k}}(h=\overline{h}-0.1)$
and its integrated value around the M-M line $I_{\rm line}(\overline{h}) =\sum_{\bm{k} \in \rm M-M\ line} \Delta n_{\bm{k}}(\overline{h})$.
Here $\Delta n_{\bm{k}}$ represents the quantity of the JW fermions introduced by an increment in $h$. They 
can be interpreted as the low-energy excitations in the Majorana spectrum, 
because the Majorana dispersion is a hybridization 
between the spectrum of the complex fermions and their particle-hole-symmetrized counterparts.
Figure \ref{nk} (b) shows that $\Delta n_{\bm{k}}$ exhibits peaks in 
the K-$\Gamma$ line, which indicates the existence of Majorana cones.
With increasing $h$, the positions of the peaks move to $\Gamma$ point.
At $\overline{h}=h_{c1}$, the line structure of $\Delta n_{\bm{k}}$ appears along the M-M line, 
which indicates the formation of the nodal line of the Majorana fermions.
This is also suggested by the enhancement of the integrated value $I_{\rm line}$ at $h_{c1}$ in Fig. \ref{nk} (c).
The behavior of the Majorana spectrum expected from $\Delta n_{\bm{k}}$ can be summarized as follows:
The Dirac point of the Majorana fermions moves along the 
K-$\Gamma$ line and it changes to the nodal line at the 
transition point ($h=h_{c1}$). Above $h_{c1}$,
the Dirac point of the Majoran fermions appears again and they move to $\Gamma$ point.
These behaviors are consistent with results obtained by 
the MF approximation~\cite{Nasu2018,Liang2018}.
We note that the line formation is not 
found clearly below or above $h_{c1}$ 
[see the left and right panels in Fig. \ref{nk} (b)].

%
\section{Ferromagnetic case}
\begin{figure}[b]
  \begin{center}
   \includegraphics[width=80mm]{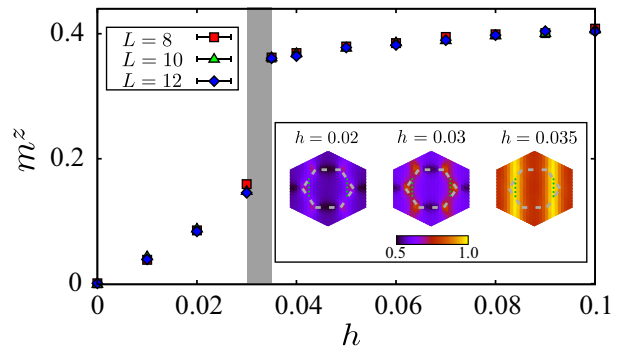}
  \end{center}
  \caption{(Color online) Magnetization process in the ferromagnetic Kitaev model.
  The inset shows the momentum distribution $n_{\bm{k}}$ for $L=12$.
  The gray bar marks the transition point.
  }
  \label{ferro}
\end{figure}

%
Finally, we show the results for the ferromagnetic coupling $K < 0$.
Figure \ref{ferro} shows the magnetization process for the ferromagnetic Kitaev model.
There clearly is only one first order transition in the magnetization curve, 
which is consistent with the previous studies\cite{Nasu2018,Liang2018}.
As is true for the antiferromagnetic case, 
the critical field $h_c\sim0.0325$ is weaker than that 
of the previous MF result $(h_c^{\rm MF} \sim 0.042)$\cite{Nasu2018}.

The momentum distributions at each $h$ are shown in the inset in Fig. \ref{ferro}.
For a weak field, $h < h_c$, there are six points 
with strong intensities around the Majorana cones, which is the same as in the antiferromagnetic case.
However, the $h$-dependence of $n_{\bm{k}}$ for $K<0$ is 
significantly different from that for $K>0$: 
Even for a weak field, $h < h_c$, island formation does not occur and 
the JW fermions are distributed over the entire BZ, especially around the M-M line.
This can be understood as follows: Because the ferromagnetic coupling in 
$z$-bond is parallel to the
magnetic filed $h$, 
the JW fermions line up along the $z$-bond in the real space---M-M line in the momentum space---to gain the energy.
For a strong field $h > h_c$, $n_{\bm{k}}$ is similar to that for the antiferromagnetic case above $h>h_{c2}$
because this phase is also the partially polarized state; it 
connects directly to the fully polarized state in the strong coupling limit.

%
%
%
\section{Summary}
%
%
In summary, we performed VMC calculations for the ground state of the Kitaev 
model in a [001] magnetic field in order to clarify the existence of the intermediate 
state 
reported in the previous studies\cite{Nasu2018,Liang2018}. 
We found that the emergence of the intermediate state 
appears even when we take into account the many-body correlations
beyond the MF approximation.
By analyzing the momentum distributions of the JW fermions, 
we showed that a Lifshitz-like transition 
of Majoran spectrum occurs at $h_{c1}$.
We also showed that a single first-order phase transition occurs 
for the ferromagnetic Kitaev coupling.
These results are consistent with 
those obtained by the MF approximation, which demonstrates
the qualitative validity of the MF approximation for the interacting Majorana fermions.
Our VMC results also show that fermionization of the localized spin different from the spinon representation is 
a useful and accurate way to analyze the Kitaev model in a magnetic field.
Using other types of the fermionizations\cite{Tsvelik1992,Kitaev2006,Fu2018},
highly accurate analyses would be possible within the framework of VMC for the Kitaev model with additional interactions---such as the Heisenberg term 
and the $\Gamma$ term\cite{Yamaji2014,Kim2015,Yadav2016,Katukuri2016,Winter2016}---as well as for tilted magnetic fields. 
Our study offers a firm basis for such extended treatments.

%
{\it Acknowledgments.}---
%
Our VMC code has been developed based on open-source software ``mVMC"\cite{Misawa2017}.
To obtain the exact diagonalization results, 
we used ``$\mathcal{H} \Phi$" package\cite{Kawamura2017}. 
We acknowledge Yukitoshi Motome, Joji Nasu, Yasuyuki Kato, 
and Yoshitomo Kamiya for fruitful discussion and important suggestions.
A part of calculations is done by using
the Supercomputer Center, the Institute for Solid State Physics, the University of Tokyo.
This work was supported in part by MEXT as a 
social and scientific priority issue 
(Creation of new functional devices and high-performance 
materials to support next-generation industries; CDMSI) to be tackled by 
using the post-K computer.
TM and KI were supported by Building of Consortia for the Development of 
Human Resources in Science and Technology from the MEXT of Japan. 
This work was also supported by a Grant-in-Aid for Scientific Research 
(Nos. 16K17746, 16H06345, 19K03739, 19K14645)
from Ministry of Education, Culture, Sports, Science and Technology, Japan.  
\bibliographystyle{prsty}
\bibliography{reference}

\end{document}